\documentclass[twocolumn,showpacs,showkeys,pra]{revtex4}
\usepackage{graphicx}%
\usepackage{dcolumn}
\usepackage{amsmath}
\usepackage{latexsym}

\begin{document}

\def\f{\frac}
\def\rdown{\rho_{\downarrow}}
\def\pa{\partial}
\def\th{\theta}
\def\Ga{\Gamma}
\def\ka{\kappa}
\def\bea{\begin{eqnarray}}
\def\eea{\end{eqnarray}}
\def\be{\begin{equation}}
\def\ee{\end{equation}}
\def\pa{\partial}
\def\d{\delta}
\def\K{\kappa}
\def\a{\alpha}
\def\eps{\epsilon}
\def\th{\theta}
\def\na{\nabla}
\def\nn{\nonumber}
\def\lan{\langle}
\def\ran{\rangle}
\def\pr{\prime}
\def\rarrow{\rightarrow} 
\def\larrow{\leftarrow}

\draft
\title{Hartman effect and non-locality in quantum networks}
\author{Swarnali Bandopadhyay}
\affiliation{S. N. Bose National Centre for Basic Sciences, JD Block,
Sector III, Salt Lake City, Kolkata 700098, India}
\email{swarnali@bose.res.in} 
\author{A. M. Jayannavar}
\email{jayan@iopb.res.in}  
\affiliation{Institute of Physics, Sachivalaya Marg, 
Bhubaneswar 751005, India} 

\date{\today}

\begin{abstract}                          
We study the phase time for various quantum mechanical networks having 
potential barriers in its arms to find the generic presence of Hartman effect. 
In such systems it is possible to control the `super arrival'
time in one of the arms by changing  parameters on another, spatially separated 
from it. This is yet another quantum nonlocal effect. Negative time 
delays (time advancement) and `ultra Hartman effect' with negative 
saturation times have been observed in some parameter regimes.
\end{abstract}
\vskip.5cm

\maketitle
 \section{Introduction}
\label{s1}
Quantum tunneling, where a particle has finite probability to penetrate a
classically forbidden region is an important feature of wave mechanics.
Invention of the tunnel diode \cite{esaki}, the scanning tunneling microscope
\cite{bining} etc. make it useful from a technological point of view. In 1932
MacColl \cite{mac} pointed out that tunneling is not only characterized
by a tunneling probability but also by a time the tunneling particle
takes to traverse the barrier. There is considerable interest on
the question of time spent by a particle in a given region of space
\cite{landauer,hauge,anatha}. The recent development of nanotechnology 
brought new urgency to study the tunneling time as it is directly related
to the maximum attainable speed of nanoscale electronic devices. In a number 
of numerical \cite{num}, 
experimental \cite{steinberg,expt} and analytical study of quantum tunneling 
processes, various definitions of tunneling times have been investigated.
These different time scales are based on various different operational
definitions and physical interpretations.
Till date there is no clear consensus about the existence of a simple 
expression for this time as there is no hermitian operator associated with it 
\cite{landauer}. 
Among the various time scales, `dwell time' \cite{smith} which gives the 
duration of a particle's stay in the barrier region regardless of how it 
escapes can be calculated as the total probability of the particle inside the
barrier divided by the incident probability current. B\"uttiker and
Landauer proposed \cite{BLtime} that one should study `tunneling time' using
the transmission coefficient through a static barrier of interest, 
supplemented by a small oscillatory perturbation. A large number of researchers
interpret the phase delay time \cite{hauge,wigner,hartman} as the temporal 
delay of a transmitted wave packet. This time is usually taken as the difference
between the time at which the peak of the transmitted packet leaves the
barrier and the time at which the peak of the incident Gaussian wave packet 
arrives at the barrier. Within the stationary phase approximation 
the phase time can be calculated 
from the energy derivative of the `phase shift' in the transmitted or 
reflected  amplitudes. B\"uttiker-Landauer \cite{BLtime} raised objection 
that the peak is not a reliable characteristic of packets distorted
during the tunneling process. 
In contrast to `dwell time' which can be defined locally, the
`phase time' is essentially asymptotic in character \cite{haugeflack}. 
The `phase time' statistics is intimately connected with dynamic
admittance of micro-structures \cite{buttiker}. This `phase time' is
also directly related to the density of states \cite{buttiker2}. 
The universality of `phase time' distributions in random and chaotic
systems has already been established earlier \cite{jayan}. In the case of `not
too opaque' barriers, the tunneling time evaluated either as a simple 
`phase time' \cite{hauge} or calculated through the 
analysis of the wave packet behaviour \cite{recami} becomes independent of the
barrier width. This phenomenon is termed as the Hartman effect 
\cite{hartman,recami,fletcher}. This implies that for sufficiently large 
barriers the effective velocity of the particle can become arbitrarily large, 
even larger than the light speed in the vacuum (superluminal effect). Though
this interpretation is a little far fetched for non-relativistic Schr\"dinger 
equation as velocity of light plays no role in it, this effect
has been established even in relativistic quantum mechanics.

Though experiments with electrons for verifying 
this prediction are yet to be done, the formal identity between 
the Schr\"odinger equation and 
the Helmholtz equation for electromagnetic wave enables one to 
correlate the results for electromagnetic and microwaves 
to that for electrons. Photonic experiments show 
that electromagnetic pulses travel with group velocities in excess of the 
speed of light in vacuum as they tunnel through a constriction in a
waveguide ~\cite{nimtz}. Experiments with photonic band-gap structures 
clearly demonstrate that `tunneling photons' indeed 
travel with superluminal group 
velocities ~\cite{steinberg}. Their measured tunneling time is practically 
obtained by comparing the two peaks of the incident and transmitted wave 
packets. Thus all these experiments directly or indirectly
confirmed the occurrence of Hartman effect without violating  
`Einstein causality' {\em i.e.}, the signal velocity or the information 
transfer velocity is always bounded by the velocity of light. 
This tunneling time could be interpreted as the `time of passage' of the
peak. Since the velocity of the `peak' may exceed arbitrarily large 
numbers, this `fast tunneling' has been frequently related to `superluminal
propagation' \cite{nimtz}. The `Hartman effect' has been extensively 
studied both for nonrelativistic (Schr\"odinger equation) and relativistic 
(Dirac equation) \cite{landauer,hauge,nimtz} cases. Recently Winful 
\cite{winful03} showed that the saturation of phase time is a direct 
consequence of saturation of integrated probability density under the barrier.
Due to this saturation addition of a new particle in the incident side leads to
an almost immediate release of another particle from the other side and these 
two particles are causally unrelated; {\em i.e.} even the superluminal  
tunneling does not violate causality. Equivalently, for electromagnetic waves 
the origin of the Hartman effect has been traced to stored energy. Since the 
stored energy in the evanescent field decreases exponentially within 
the barrier after a certain decay distances it becomes independent of the 
width of the barrier. The Hartman effect has been found in 
one dimensional barrier tunneling \cite{recami} as well as for 
cases beyond one dimension as in tunneling through mesoscopic rings 
in presence of Aharonov-Bohm flux \cite{swarnali-AB}. In the current note 
we extend the study of phase times for branched networks of quantum wires.
 Such networks can readily be realized in optical wave propagation 
experiments. 

In Section-\ref{description} we give the description of  typical 
quantum network systems under consideration. The theoretical framework used to 
analyze these systems is provided in Section-\ref{theory}. All the main 
results showing Hartman effect and the effects of quantum non-locality 
are discussed in Section-\ref{result}. Finally we summarize the results and 
draw conclusions in Section-\ref{conclusion}.

\section{Description of the system}
\label{description}
\begin{figure}[t]
\begin{center}
\includegraphics[width=8.0cm]{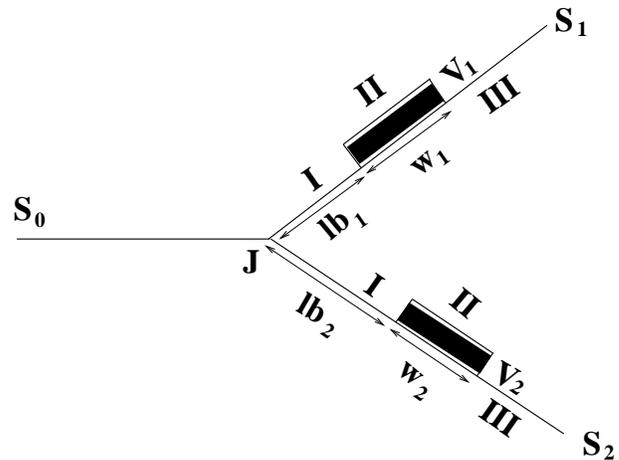}
\end{center}
\caption{Schematic diagram of a Y-junction or three-way splitter.} 
\label{system}
\vskip .5cm
\end{figure}

As a model system, we choose a network of 
thin wires. The width of these
wires are so narrow that only the motion along the length of the wires
is of interest (a single channel case). The motion in the perpendicular 
direction is frozen in the
lowest transverse subband. In a three-port Y-branch circuit 
(Fig.~\ref{system}) two side branches of quantum wire $S_1$ and $S_2$ are 
connected to a `base' arm $S_0$ at the
junction $J$. In general one can have $N (\geq 2) $ such
side branches connected to a `base' wire.

We study the scattering problem across a network geometry as
 presented schematically in Fig.~\ref{system}. 

Such geometries are important from the point of view of basic science due to  
their properties of tunneling and interference~\cite{webb,gefen} 
as well as in applications 
such as wiring in nano-structures. In particular, the Y-junction carbon 
nanotubes are in 
extensive studies and they show various interesting properties like 
asymmetric current voltage characteristics~\cite{cnr}. 
In our system of interest there are finite 
quantum mechanical potential barriers of strength 
$V_n$ and width $w_n$ in the $n$-th side branch.
The number of side branches can vary from $n=2,3,\cdots N$. 
We focus on a situation wherein 
the incident electrons have an energy $E$ less than $V_n$ for all $n$. 
The impinging electrons in this subbarrier 
regime travels as an evanescent mode/wave and the transmission involve 
contributions from quantum tunneling and multiple
reflections between each pair of barriers and the junction point. 
Here we are interested in a single channel case where the 
Fermi energy lie in the lowest subband. To excite the evanescent modes 
in the side branches one has to produce constrictions by making the 
width of the regions of wires containing barriers much thinner than that of 
other parts of the wires. The electrons 
occupying the lowest subband in the connecting wire on entering the 
constrictions 
experience a potential barrier (due to higher quantum zero point energy)
and propagate as an evanescent mode \cite{amj}.
In this work an analysis of the phase time or 
the group delay time in such a system is carried out.

\section{Theoretical treatment}
\label{theory}   
We approach this scattering problem using the quantum wave guide theory 
~\cite{xia,jayan2}. In the stationary case the incoming particles
are represented by a plane wave $e^{ikx}$ of unit amplitude. The
effective mass of the propagating particle is $m$ and the energy is 
$E=\hbar^2 k^2/2m$ where $k$ is the wave vector corresponding to the free 
particle.
The wave functions, which are solutions of the Schr\"odinger equation, 
in different regions  
of the system considered (Fig.~\ref{system}) can be written as,

\begin{widetext}
\begin{eqnarray}
\psi_{in}(x_0) &=& e^{ikx_0} + R e^{-ikx_0}\, \, \, \,
 (\mbox{in}\, S_0) \, ,\nn\\
\psi_{(n)_{I}}(x_n) &=& A_n\, e^{i\,k\,x_n} + B_n\, e^{-i\,k\,x_n}\, \, \, \, 
(\mbox{region I in}\, S_n) \, , \nn\\
\psi_{(n)_{II}}(x_n) &=& C_n\, e^{-\kappa_n \,(x_n-lb_n)} + D_n\, e^{\kappa_n \,(x_n-lb_n)} \, \, \, \, (\mbox{region II in}\, S_n) \,,\nn\\
\psi_{(n)_{III}}(x_n)&=& t_n\, e^{i\,k\,(x_n-lb_n-w_n)} \, \, \, \,
(\mbox{region III in}\, S_n) \,,\nn
\end{eqnarray}
\end{widetext}

with $\kappa_n=\sqrt{2m(V_n-E)/\hbar^2}$ being the   
imaginary wave vector in presence of rectangular barrier of
strength $V_n$. $\psi_{(n)_{I}}, \psi_{(n)_{II}}$ and $\psi_{(n)_{III}}$ 
denote wave functions in three regions $I, II$ and $III$, respectively, 
on $n$-th side branch. 
$x_0$ is the spatial coordinate for the 
`base' wire, whereas $x_n$ is spatial coordinate for the $n$-th  
arm. All these coordinates are measured from the junction $J$. 
In $n$-th side branch, the barrier starts at a distance $lb_n$ from the 
junction $J$.
 
We use Griffith's boundary conditions~\cite{griffith}
\begin{equation}
\psi_{in}(J)=\psi_{(n=1)_{I}}(J)=\psi_{(n=2)_{I}}(J)=\cdots =\, \psi_{(n=N)_{I}}(J),\label{bc1}
\end{equation}
and 
\begin{equation}
\frac{\partial \psi_{in}(x_0)}{\partial x_0}\Big |_J = 
\Sigma_n \frac{\partial \psi_{(n)_{I}}}{\partial x_n}\Big |_J \, ,
\label{bc2}
\end{equation}
at the junction $J$. All the derivatives are taken either outward or inward 
from the junction~\cite{xia}. In each side branch, at the starting and end 
points of the barrier, the boundary conditions
can be written as
\begin{eqnarray}
\psi_{(n)_{I}}(lb_n)=\psi_{(n)_{II}}(lb_n)\, ,\label{bc3} \\
\psi_{(n)_{II}}(lb_n+w_n)=\psi_{(n)_{III}}(lb_n+w_n)\, ,\label{bc4} \\
\frac{\pa \psi_{(n)_{I}}}{\pa x_n}\Big |_{(lb_n)} = 
\frac{\pa \psi_{(n)_{II}}}{\pa x_n}\Big |_{(lb_n)} \, ,\label{bc5} \\
\frac{\pa \psi_{(n)_{II}}}{\pa x_n}\Big |_{(lb_n+w_n)} = 
\frac{\pa \psi_{(n)_{III}}}{\pa x_n}\Big |_{(lb_n+w_n)} \, .\label{bc5} 
\end{eqnarray} 

From the above mentioned boundary conditions one can obtain 
the complex transmission amplitudes $t_n$ on each of the side branches.

\section{Results and Discussions}
\label{result}
Once $t_n$ is known, the `phase time' (phase time for transmission) can be 
calculated from the energy derivative of the phase of the transmission 
amplitude \cite{hauge,wigner} as
\begin{equation}
\tau_n = \hbar \, \frac{\partial Arg[t_n]}{\partial E}\, , 
\label{phtm}
\end{equation}
where, $v = \hbar k/m$ is the velocity of the free particle. 
The concept of `phase delay time'
was first introduced by Wigner \cite{wigner} to estimate how long a 
quantum mechanical wave packet is delayed by the scattering obstacle.

\begin{figure}[b]
\vskip 1cm
\begin{center}
\includegraphics[width=9.0cm]{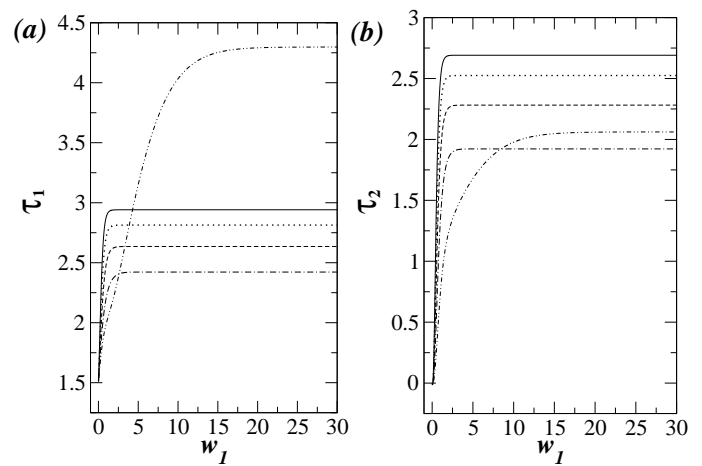}
\end{center}
\caption{For a 3-way splitter with a barrier in $S_1$ arm, the `phase times' $\tau_1$ and $\tau_2$ are plotted as a function of barrier width `$w_1$' in $(a)$ and $(b)$ respectively. The solid, dotted, dashed, dot-dashed and the dashed-double dotted curves are for $V_1=5, 4, 3, 2,$ and $1.05$ respectively. Other system parameters are $ E = 1, lb_1 = 3$.} 
\label{1barrier}
\end{figure}

In what follows, let us set $\hbar =1$ and $2m =1$. We now proceed to analyze
the behavior of $\tau_n$ as a function of various 
physical parameters for different network topologies. We measure
time at the far end of each barrier in the branched
arms containing barriers and in the case of arms in absence of any barrier
we measure the phase time at the junction points. We express all the
physical quantities in dimensionless units {\em i.e.} all the barrier strengths 
$V_n$ in units of incident energy $E$ ($V_n \equiv V_n/E$), all the
barrier widths $w_n$ in units of inverse wave vector 
$k^{-1}$ ($w_n \equiv k w_n$), where 
$k=\sqrt E$ and all the extrapolated phase time $\tau_n$ in units of inverse 
of incident energy $E$ ($\tau_n \equiv E\tau_n$).

First we take up a system similar to the Y-junction shown in 
Fig.~\ref{system} in presence of a 
barrier $V_1$ of width $w_1$ in arm $S_1$ but in absence of any barrier
in arm $S_2$. For a tunneling particle having energy $E < V_1$ we find
out the phase time $\tau_1$ in arm $S_1$ as well as $\tau_2$ in arm $S_2$ 
as a function of barrier width $w_1$ (Fig.~\ref{1barrier}).
From Fig.~\ref{1barrier}(a) it is clear that $\tau_1$ evolves with $w_1$ and 
eventually saturates to $\tau_{s1}$ for large $w_1$ to show the Hartman 
effect. Fig.~\ref{1barrier}(b) shows the phase time $\tau_2$ in arm $S_2$ 
which does not contain any barrier. This also evolves and saturates with 
$w_1$, the length of the barrier in the other arm $S_1$. This delay 
is due to the 
contribution from paths which undergo multiple reflection in the first
branch before entering the second branch via junction point $J$. In absence 
of a barrier in the $n$-th arm the phase 
time $\tau_n$ measured close to the junction $J$ should go 
to zero {\em i.e.} $\tau_n \to 0$ in the absence of multiple scatterings in 
the first arm. 
Note that $\tau_{s1}$ and $\tau_{s2}$ change with energies of the incident
particle (Fig.~\ref{1barrier}). From Fig.\ref{1barrier} it can be easily
seen that $\tau_{s2}$ is always smaller than $\tau_{s1}$ for any particular 
$V_1$ {\em i.e.} the saturation time in the arm having no barrier is smaller. 
The phase time in both the arms show non-monotonic behavior as a function of
$V_1$. As we decrease the strength of the barrier $V_1$ the value 
of $\tau_1$ ($\tau_2$) decreases in the whole range of widths of the barrier 
and also the saturated value of $\tau_{s1}$ ($\tau_{s2}$) decreases until 
$V_1$ reaches $1.6$ and with further decrease in $V_1$ the values of 
$\tau_1$ ($\tau_2$) as well as $\tau_{s1}$ ($\tau_{s2}$) starts increasing.

\begin{figure}[t]
\begin{center}
\includegraphics[width=8.0cm]{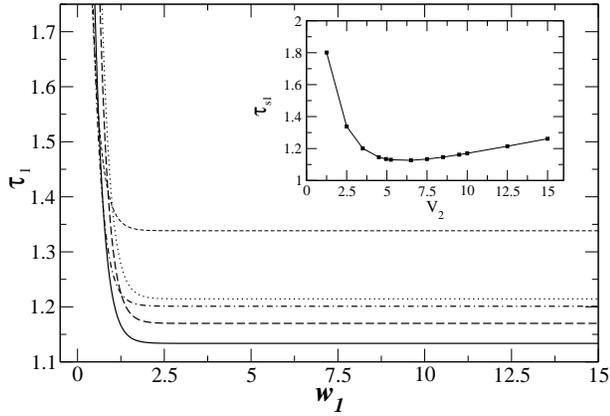}
\end{center}
\caption{Here for a 3-way splitter with one barrier in each branched 
arm $S_1$ and $S_2$, the `phase times' $\tau_1$ is plotted as a function 
of barrier width `$w_1$' keeping $w_2 (=1)$ and $V_1 (=5)$ fixed and for 
different values of parameter $V_2$. The 
small dashed, dot-dashed, solid, long dashed and dotted curves are 
for $V_2 = 2.5, 3.5, 5.0, 10.0$ and $12.5$ 
respectively. Other system parameters are $E=1, lb_1 = lb_2 = 3$. In the
inset $\tau_{s1}$ is plotted as a function of $V_2$ for the same system
parameters.} 
\label{nonlocal-hartman}
\vskip .5cm
\end{figure}

As the second case we take up another Y-junction which contain potential 
barriers in both its side branches as shown in Fig.~\ref{system}. We fix the
values of $V_1 (=5)$ and vary $w_1$ for each values of $V_2$
to study the $w_1$-dependence of $\tau_1$ (Fig.\ref{nonlocal-hartman}). 
From Fig.~\ref{nonlocal-hartman} we see that $\tau_1$ decreases 
with increase in $w_1$ to saturate to a value $\tau_{s1}$ at each value of 
$V_2$ thereby showing `Hartman effect' for arm `$S_1$'.
But now, we can tune the saturation phase time at arm $S_1$ non-locally 
by tuning strength of the barrier potential $V_2$ sitting on another arm 
$S_2$! 
{\it Thus `quantum nonlocality' enables us to control the `super arrival' time
in one of the arms ($S_1$) by changing a parameter ($V_2$) on another, 
spatially separated from it}. In the inset of Fig.~\ref{nonlocal-hartman} we
plot $\tau_{s1}$ as a function of $V_2$. It clearly shows that when 
the barrier strengths $V_1$ and $V_2$ are very close the `phase time' 
reaches 
its minimum value. In all other cases {\em i.e.} whenever $V_1 \ne V_2$, the 
value of $\tau_{s1}$ is larger.     

\begin{figure}[t]
\begin{center}
\includegraphics[width=8.0cm]{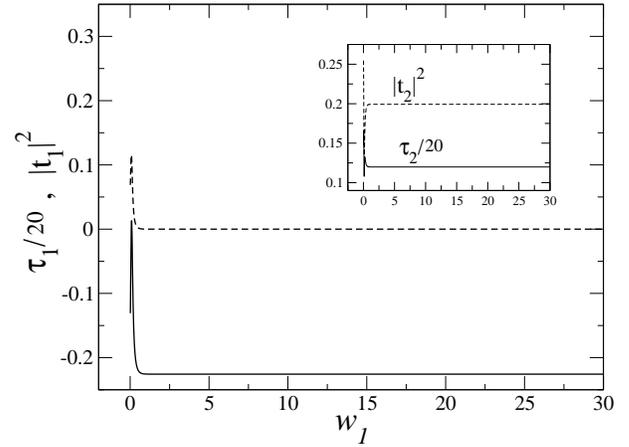}
\end{center}
\caption{Here for a 3-way splitter with one barrier in each side branch $S_1$
 and $S_2$, the `phase times' $\tau_1$ (solid curve) and $|t_1|^2$ (dashed)
are plotted as a function of `$w_1$' for a very small 
$w_2 (=0.5)$. Other system parameters are $E=1$, $lb_1=lb_2=2.5$, $V_2=5$ 
and $V_1=15$. In the inset, the solid and dashed 
curves represent $\tau_2$ and $|t_2|^2$ respectively as a function of $w_1$. For better visibility we have plotted phase times scaled down by a factor of $20$.} 
\label{2bar-negative}
\vskip .5cm
\end{figure}

We will show now another interesting result related to the Hartman effect. 
For this we keep $V_2 (=5)$ unaltered and reduce $w_2$.  
For very small $w_2 (=0.5)$ we see from Fig.~\ref{2bar-negative} that 
$\tau_1$ is negative for almost the whole range of $w_1$-values
 showing `time-advancement' and eventually after a sharp decrease saturates 
to a negative value of $\tau_{s1} =  -4.514$  
implying `Hartman effect' with advanced time.
It might be noted that, in principle, the `time-advancement' 
(Fig.~\ref{2bar-negative}) can be measured
experimentally as $|t_1|^2$ has a non-zero finite value for a small 
range of $w_1$ at lower $w_1$ regime where $\tau_1$ is negative. 
In the inset we plot the corresponding $\tau_2$ and $|t_2|^2$ as function 
of $w_1$. Again the values of $\tau_2$ remains different from the one 
dimensional tunneling through a barrier of strength $V_2$ and width $w_2$
in the whole range of $w_1$ implying `quantum nonlocality'. In the
cases discussed so far $\tau_2$ vary more sharply in small $w_1$
regime. Further the inset in Fig.~\ref{2bar-negative} shows a dip in 
$\tau_2$ at parameter regimes where $|t_2|^2$ has a minimum. For a wave packet 
with large spread in real space it is 
possible that the leading edge of the wave packet reaches the barrier much
earlier than the peak of the packet. This leading edge in turn can tunnel 
through to produce a peak in the other end of the barrier much before the 
incident wave packet reaches the barrier region, sometimes referred to as pulse 
reshaping effect. This, in general, causes `time advancement' \cite{landauer}.
This negative delay does not violate causality, however, the time is bounded 
from the below. In the presence of square wells in one dimensional systems 
negative time delays have been observed. This effect is termed as 
`ultra Hartman effect' [ see for details \cite{muga02} ].

\begin{figure}[t]
\begin{center}
\includegraphics[width=8.0cm]{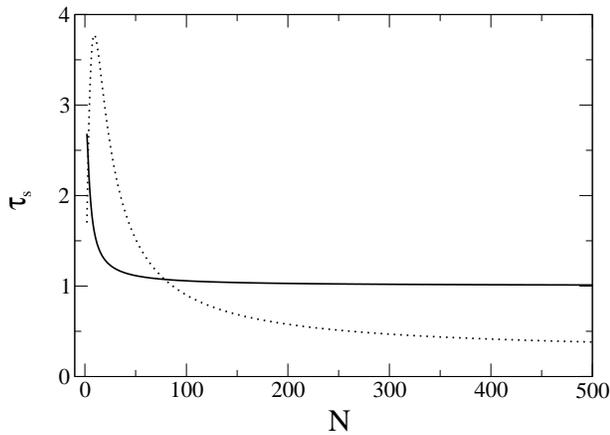}
\end{center}
\caption{Here $N (\ge 2)$ number of side branches each with a barrier 
of strength $V (=5)$ and width $w (=100)$ are connected with the
incident arm $S_0$. In each arm barriers start at the same point $lb (=1)$. 
Thus all the side branches being identical the `phase time'
for transmission through these arms $S_n, n=2,3,\cdots N$ saturate to
the same value $\tau_s$. The saturated `phase time' $\tau_s$ is plotted as a 
function of the total number of side branches $N$ in the system. The dotted 
and solid curves are for $V = 5.0$ and $1.25$ respectively. The incident energy is kept at $E=1$.} 
\label{tauS-N}
\vskip .5cm
\end{figure}

Finally consider a similar system as that shown in
Fig.~\ref{system}, but in presence of $N (\ge 2)$ identical side branches
and study phase time as a function of increasing $N$. 
All the side branches being 
identical the `phase times' for transmission through each of these arms 
$S_n, n=2,3,\cdots N$ saturate to the same value $\tau_s$ for very
large $w_n$. In Fig.\ref{tauS-N}
we plot the saturation value $\tau_s$ as a function of the total number 
of side branches $N$ present in the system. From the figure we see that
for $V = 5$, $\tau_s$ first increases with $N$ to a maximum value of 
$3.776$ at $N=9$ and thereafter keeps on decreasing with the increase 
of $N$. As we start reducing the
strength of the barriers from $5$ we see that for $V = 1.49$ the 
increasing nature of $\tau_s$ in small $N$ range vanishes. In 
general, at larger $N$, the decreasing nature of $\tau_s$ with $N$ persists, 
{\em e.g.}, note the solid curve in Fig.\ref{tauS-N} plotted at $V = 1.25$, 
but the initial increase in $\tau_s$ is not a generic feature. For larger $N$
transmission amplitude in each side branch reduces with increase in $N$
and hence the corresponding peaks of wave packets reach the far end at earlier 
times thereby reducing $\tau_s$.  

\section{Conclusions}
\label{conclusion}
We have studied Hartman effect and non-locality in quantum network consisting 
of a main one dimensional arm having $N (\ge 2)$ side branches. 
These side branches may or may not have barriers. In presence of barrier the 
`phase time' for  transmission through a side branch shows the 
`Hartman effect'. 
In general, as the number of side branches $N$ increases, the saturated 
`phase time' decreases. Due to quantum nonlocality the `phase time' and it's 
saturated value at any side branch feels  the presence of barriers in other 
branches. Thus one can tune the saturation value of
`phase time' and consequently the superluminal speed in one branch 
by changing barrier strength or width in any other 
branch, spatially separated from the former. 
Moreover Hartman effect with negative saturation times (time advancement) 
has been observed for some cases. In conclusion generalization of Hartman
effect in branched networks exhibits several counter-intuitive results 
due to quantum non-locality. System parameters such as number of side branches 
$N$ and barrier widths $w_n$, strengths $V_n$, distance $lb_n$ (from $J$)
and incident energy $E$ etc. play very sensitive roles in determining delay 
times. The delay times are also sensitive to the junction $S$-matrix
elements used for a given problem. In our present problem junction 
$S$-matrix is determined uniquely by the wave guide transport methods. 
Depending on $lb_n$ there may be one or several bound states located 
between the barriers in different branched arms and as a consequence
saturated delay time can be varied from the negative (ultra Hartman
effect) to positive and vice-versa. We have verified this by looking at the
transmission coefficient in the second arm $S_2$ which exhibits clear resonances
as a function of $lb_2$. Around these resonances, saturated delay time in
the first arm $\tau_{s1}$ changes the sign [the details of which will be
published elsewhere \cite{swarnali-nu}]. Moreover the reported effects are 
amenable to experimental verifications in the electromagnetic wave-guide 
networks.

\section{Acknowledgments}
One of the authors (SB) thanks Debasish Chaudhuri and Prof. Binayak Dutta-Roy
for several useful discussions.

\end{document}